\documentclass[aps,prl,groupedaddress,amsmath,amssymb,twocolumn]{revtex4-1}
\usepackage{graphicx,float} 
\usepackage{braket}
\usepackage[utf8]{inputenc}
\usepackage{dcolumn}
\usepackage{xfrac}

\newcommand{\gras}[1]{\boldsymbol{#1}}
\newcommand{\di}{i} % default math "i"
\newcommand{\phin}{\varphi_{n}}
\newcommand{\phip}{\varphi_{p}}
\newcommand{\angles}{\mathbf{x}}
\newcommand{\qvec}{\gras{q}}

\begin{document}

\title{Fission of $^{240}$Pu with Symmetry-Restored Density Functional Theory}

\author{P. Marevi\'c}
\affiliation{Nuclear and Chemical Sciences Division, Lawrence Livermore National Laboratory, Livermore, CA
94551, USA}
\author{N. Schunck}
\affiliation{Nuclear and Chemical Sciences Division, Lawrence Livermore National Laboratory, Livermore, CA
94551, USA}

\begin{abstract}  
Nuclear fission plays an important role in fundamental and applied science, 
from astrophysics to nuclear engineering, yet it remains a major challenge to 
nuclear theory. Theoretical methods used so far to compute fission 
observables rely on symmetry-breaking schemes where basic information on the 
number of particles, angular momentum, and parity of the fissioning nucleus is lost. In 
this letter, we analyze the impact of restoring broken symmetries in the 
benchmark case of $^{240}$Pu.
\end{abstract}
 
\date{\today}

\maketitle

%
% NOTES
%  - different fission channels => different initial conditions of spin/parity for the nucleus
%  - characteristics of fission fragments (Z, N, spin, E*) essential to model their decay 
%

% Physics of fission
{\it Introduction.} Nuclear fission, the process of splitting an atomic nucleus 
into two or more fragments, is a key ingredient for modeling
%stellar (NS: not stellar, r-process only)
nucleosynthesis, as it prevents the production of superheavy elements and its 
products (fission fragments, neutrons, and photons) impact the astrophysical 
reaction rates \cite{beun2008,mumpower2016,mumpower2018,mumpower2019,vassh2019}. 
Detailed knowledge of the fission channels (spontaneous, neutron-induced, etc.) of selected actinide nuclei is also 
at the heart of important societal applications in medicine, energy production, 
nuclear forensics and safeguards, etc. \cite{hayes2017}. In spite of recent 
advances in experimental techniques \cite{andreyev2017,schmidt2018}, 
measurements are not always possible and theoretical simulations that model the 
entire process leading to the formation and decay of fission fragments are 
mandatory. 

% DFT description of fission
Despite formidable efforts over the past eighty years, a fully microscopic 
description of the fission phenomenon based on nuclear forces among protons and 
neutrons and quantum many-body methods remains a challenge \cite{krappe2012,
schunck2016}. Nuclear density functional theory (DFT) is currently the only 
fully quantum-mechanical framework that can be used to compute
fission observables, such as spontaneous fission half-lives \cite{baran2015} or fission fragment 
distributions \cite{goutte2004,regnier2016a,tao2017,regnier2019,verriere2019,
zhao2019,zhao2019a}. Time-dependent DFT provides a natural framework to 
explain the energy sharing among the fragments \cite{simenel2014,bulgac2016,bulgac2019}, 
which is key to predicting their deexcitation. 

% DFT for FPY
Following the initial insight of Bohr and Wheeler, most DFT-based approaches to 
fission are built upon the assumption that a small set of collective degrees of 
freedom (typically related to the deformation of the nuclear shape) drives the fission 
process \cite{bohr1939,klein1991,schunck2016}. This description is formalized 
through the concept of spontaneous symmetry breaking: the
intrinsic nuclear density does not conserve symmetries of the nuclear Hamiltonian 
\cite{schunck2019,sheikh2019}. In particular, the geometrical deformation of a nucleus is manifested by the breaking of rotational, axial, or reflection symmetry; nuclear superfluidity \cite{brink2005} by the breaking of particle number symmetry, etc. The corresponding potential energy surfaces (PESs) encode the total energy as a function of order parameters associated with breaking   
each symmetry. They can be used to infer important quantities of interest, 
from tunneling probabilities for spontaneous fission half-lives 
\cite{baran2015}, to the determination of initial states for time-dependent 
approaches \cite{bulgac2016,bulgac2019}, or basis states for quantum 
configuration mixing with the generator coordinate method (GCM) 
\cite{reinhard1987,berger1991}. The most advanced PESs for fission 
studies are based on solving the Hartree-Fock-Bogoliubov (HFB) equation with 
Skyrme \cite{staszczak2009,schunck2014}, Gogny \cite{warda2012,guzman2014}, or 
relativistic \cite{lu2012,zhao2015} functionals. 

However, such approaches conceal the basic information on quantum numbers related to each broken symmetry, such as the particle number, angular momentum, and parity of a nucleus. Restoring these symmetries is especially important to model different fission 
channels. For example, symmetry-breaking theory is incapable of distinguishing 
between the neutron-induced fission of $^{235}$U and the photofission of 
$^{236}$U \cite{krishichayan2019}. In both cases, the compound nucleus is the 
same, $^{236}$U, but the spin-parity distribution can be substantially 
different since $^{235}$U($n,f$) involves coupling the $^{235}$U ground-state 
angular momentum $J = \sfrac{7}{2}$ with the spin distribution of the neutron beam, while 
$^{236}$U($\gamma,f$) couples the spin 1 of the photon with $J=0$ of an 
even-even nucleus. Symmetry restoration techniques are also essential to obtain 
more realistic estimates of fission fragment characteristics, as was shown in 
the simplest case of particle number restoration \cite{scamps2017,
verriere2019}. Finally, correlation energies induced by symmetry 
restoration modify the overall PES, which could 
impact fission dynamics.

% DFT approaches: practical problems
With the exception of a several pioneering works \cite{bender2004a,samyn2005,hao12,bernard2019}, 
there has been no attempt at examining the impact of symmetry restoration in 
the context of fission. In addition to formal difficulties with symmetry 
restoration for standard functionals \cite{sheikh2019,duguet2009}, the computational 
cost of probing a large number of extremely deformed configurations in heavy 
nuclei is prohibitively high, especially when simultaneously restoring multiple 
symmetries. In fact, symmetry-breaking PESs can usually only be computed by employing large 
harmonic oscillator (HO) bases with many incomplete shells which are not closed 
under spatial rotations and for which conventional algorithms of rotational 
symmetry restoration are inapplicable \cite{robledo1994}. 

% Goal of the paper
%In this letter, we use a technique to restore rotational symmetry in 
%incomplete bases originally proposed in \cite{robledo1994} but never tested before 
%to restore, for the first time, broken symmetries in $^{240}$Pu from the ground 
%state to scission.
In this letter, we implement for the first time the 
technique of rotational symmetry restoration in incomplete
bases originally proposed in \cite{robledo1994}, and perform the first symmetry restoration in $^{240}$Pu
from the ground state to scission.
High-performance computing capabilities enable us to quantify 
the effect of particle number, angular momentum, and parity projections on the underlying PES and on the fission fragment mass distributions.

{\it Method.} Symmetry-restored DFT is a two-step method. In the first step, we 
generated a set of axially symmetric HFB configurations with the 
HFBTHO package \cite{perez2017}, using the SkM* parameterization of the Skyrme 
energy functional \cite{bartel1982}, a mixed volume-surface contact pairing 
force \cite{dobaczewski2002}, and constraints on the values of the quadrupole 
$(q_{20})$ and octupole $(q_{30})$ moments. These quantities correspond to the 
elongation and the mass asymmetry of a nuclear shape, respectively, and 
arguably represent the most pertinent collective degrees of freedom for 
describing the fission phenomenon. The HFB equations were solved by expanding 
the solution in a deformed HO basis of
$N_{\rm max} = 31$ incomplete shells with the corresponding lowest 
$N_{\rm osc}=1100$ oscillator states included. 
The oscillator frequency and the basis deformation parameter were optimized for 
each $\qvec \equiv (q_{20}, q_{30})$ configuration separately; more details on technical aspects of the HFB calculation can be found in Ref.~\cite{schunck2014}. 

In the next step, collective correlations related to the restoration of 
symmetries were incorporated by projecting the HFB configurations onto good values of angular momenta $J $, particle 
numbers $(N, Z)$, and parity $\pi$. The projected kernels 
$\mathcal{K}^{J^{\pi}NZ}_{\qvec}$ play the central role in this procedure,
\begin{equation}
\mathcal{K}^{J^{\pi}NZ}_{\qvec} = 
\int_{\beta} \sum_{\varphi_n,\varphi_p} \mathcal{K}^{\beta, \varphi_n, \varphi_p}_{\qvec},
\label{eq:projected_kernels}
\end{equation}
where $\int_{\beta} \equiv \frac{2J+1}{2} \int_0^{\pi} \,d\beta \sin\beta 
d^{J*}_{00}(\beta)$ denotes integration over the rotational angle $\beta$ with small Wigner matrices $d^{J*}_{00}(\beta)$ as weights, while 
$\sum_{\varphi_n,\varphi_p} \equiv \sum_{l_n,l_p=1}^{N_{\varphi}} 
e^{-\di N_0 \varphi_{l_n}} e^{-\di Z_0 \varphi_{l_p}}$ denotes Fomenko sums 
\cite{fomenko1970} over gauge angles $\varphi_{l_{\tau}} = 
l_{\tau} \frac{\pi}{N_{\varphi}}$ ($\l_{\tau} = 0, ..., N_{\varphi}-1$) for 
neutrons ($\tau=n$) and protons ($\tau=p$). In our study, the projected kernel actually corresponds to the expectation value of the operator $\hat{O}$ in the symmetry-restored state. Therefore, the integrand of 
Eq.~(\ref{eq:projected_kernels}) can be written as
\begin{equation}
\mathcal{K}^{\angles}_{\qvec} 
= 
\braket{\Phi (\qvec)|\hat{O}e^{-\di \beta \hat{J}_y} e^{\di \phin \hat{N}}  e^{\di \phip \hat{Z}} \hat{P}^{\pi} |\Phi (\qvec)},
\label{eq:rotated_overlap1}
\end{equation}
where we introduced $\angles \equiv 
\{ \beta, \varphi_n, \varphi_p \}$ for compactness, $e^{-\di \beta \hat{J}_y} e^{\di \phin \hat{N}} e^{\di \phip \hat{Z}}
\equiv \hat{R}$ is the rotation operator, $\hat{P}^{\pi}$ is the parity projection operator, and $\hat{J}_{y}$, $\hat{N}$, and $\hat{Z}$ correspond to the $y$ component of the total angular momentum, the 
neutron number, and the proton number operators, respectively. The norm overlap 
kernel $\mathcal{N}^{J^{\pi}NZ}_{\qvec}$ is obtained by using 
Eqs.~\eqref{eq:projected_kernels} and \eqref{eq:rotated_overlap1} with the identity operator, $\hat{O}= \hat{1}$, 
and the Hamiltonian kernel $\mathcal{H}^{J^{\pi}NZ}_{\qvec}$ by using them
with the nuclear Hamiltonian $\hat{H}$.

When computing large-scale PESs, it is customary to improve convergence by 
truncating and adjusting at each point $\qvec$ the characteristics of the 
underlying HO basis. However, the resulting basis is not closed under spatial 
rotations. Formally, given the rotational symmetry transformation $\mathcal{T}$ 
and a single-particle basis defined by the creation and annihilation operators 
$\{c_l^{\dagger}, c_k \}$, the rotated basis $\mathcal{T}^{-1} 
\{ c_l^{\dagger}, c_k \} \mathcal{T}$ will contain states that are not present 
in the original basis. This prevents us from using conventional 
symmetry-restoring algorithms, which all assume closure of the basis under 
rotations. The elegant solution to this hurdle was proposed 25 years 
ago by Robledo \cite{robledo1994}, who reformulated the Wick theorem 
\cite{balian1969,hara1979} to encompass bases not closed under symmetry 
transformations. Based on the formalism of Ref. \cite{robledo1994}, we can
write the rotated norm overlap as
\begin{equation}
\mathcal{N}^{\angles}_{\qvec} 
= 
\sqrt{\det A^{\angles}_{\qvec}  \times \det R^{\angles}},
\label{eq:rotated_overlap2}
\end{equation}
where
\begin{equation}
A^{\angles}_{\qvec} 
= 
U^T_{\qvec} \big( R^{\angles T} \big)^{-1} U_{\qvec}^* + V_{\qvec}^T R^{\angles} V_{\qvec}^*.
\label{eq:A_matrix}
\end{equation}
Here, $U_{\qvec}$ and $V_{\qvec}$ are the Bogoliubov matrices corresponding to 
the HFB configuration $\ket{\Phi(\qvec)}$, and $R^{\angles}$ denotes the matrix 
of the rotation operator $\hat{R}$ in the HO basis \cite{nazmitdinov1996}. Note 
that in the case of a basis closed under rotations $|\det R^{\angles}| = 1 $, 
and the expression (\ref{eq:rotated_overlap2}) reduces to the conventional 
Onishi formula \cite{onishi1966}. In the symmetry-restored DFT framework, the 
Hamiltonian kernel $\mathcal{H}^{J^{\pi}NZ}_{\qvec}$ is a functional of the 
one-body, transition density $\rho^{\angles}_{\qvec}$ and pairing tensor 
$\kappa^{\angles}_{\qvec}$. When the basis is not closed under rotations, these 
read
\begin{subequations}
\begin{align}
\rho^{\angles}_{\qvec} &= R^{\angles} V_{\qvec}^* A_{\qvec}^{\angles-1} V_{\qvec}^T  ,
\label{eq:density} \\
\kappa^{\angles}_{\qvec} &= R^{\angles} V_{\qvec}^* A_{\qvec}^{\angles-1} U_{\qvec}^T,
\end{align}
\end{subequations}
where the mixed-density prescription was used
\cite{robledo2010}. The symmetry-restored energy is simply the ratio 
$\mathcal{E}^{J^{\pi}NZ}_{\qvec} = \mathcal{H}^{J^{\pi}NZ}_{\qvec} / 
\mathcal{N}^{J^{\pi}NZ}_{\qvec}$.

{\it Least-energy fission pathway.} Although distinct from the most 
probable fission path \cite{sadhukhan2013,giuliani2013}, the least-energy fission pathway provides valuable 
information about fission dynamics such as the existence and energies of 
fission barrier heights or fission isomers. These pseudodata are important 
%role in 
to predict the stability of superheavy elements or evaluate
neutron-induced fission cross sections, especially in regions of the nuclide 
chart where no experimental data is available \cite{sin2006,goriely2009}. 
The goal of the present analysis is 
to assess the effect of symmetry restoration on such data  
along the entire fission pathway. In this regard, it represents an extension of 
the early work by Bender and co-workers who studied the effect of symmetry 
restoration along the reflection-symmetric ($q_{30} = 0$) pathway and up to 
moderate deformations only \cite{bender2004a}. 

\begin{figure}[!htb]
\includegraphics[width=0.49\textwidth]{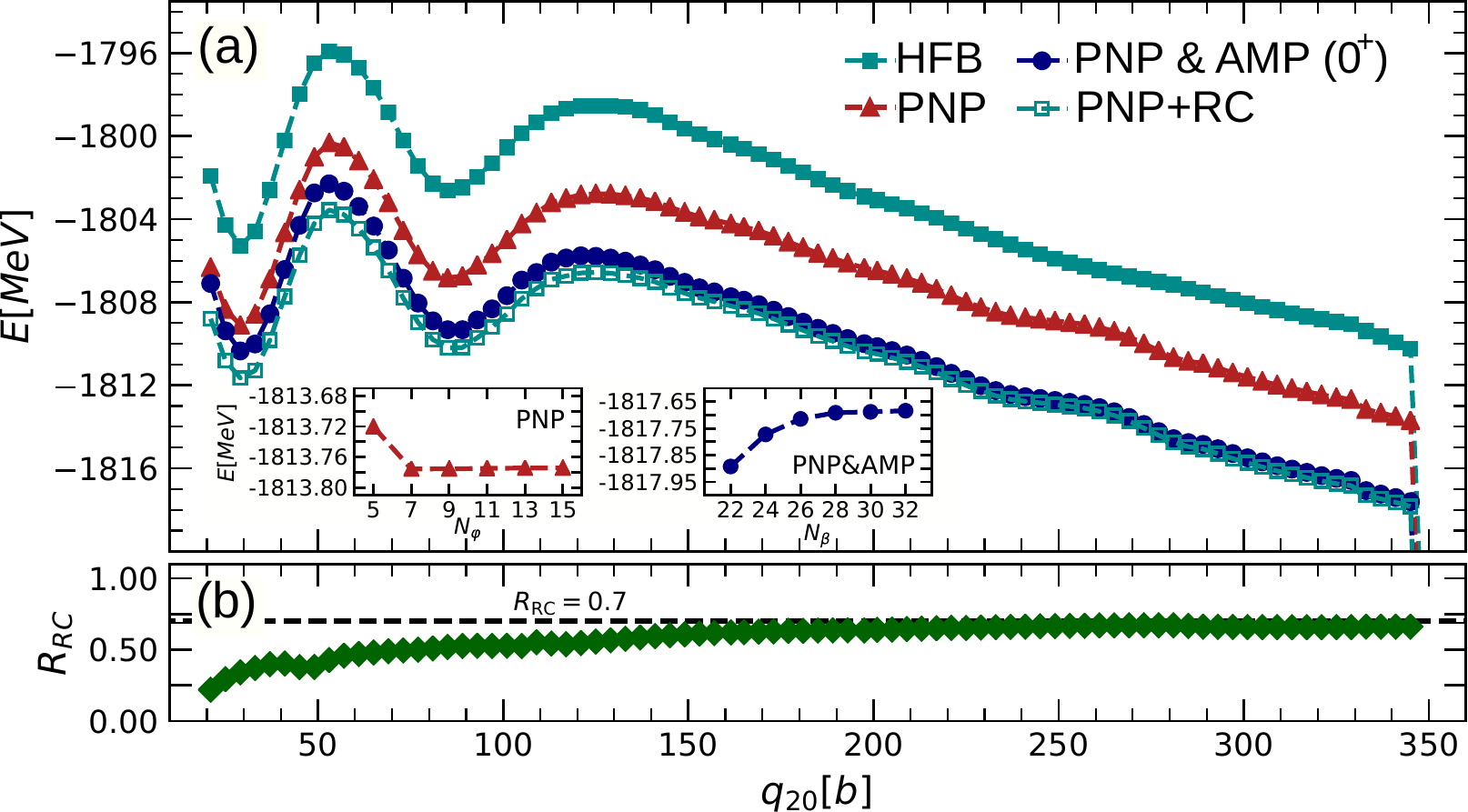}
\caption{(a): Least-energy fission pathway in $^{240}$Pu calculated 
in the HFB approximation (turquoise squares); projected onto good values of particle 
numbers (PNP, red triangles); projected onto good values of particle numbers, angular momentum 
($J=0$) and parity (PNP\&AMP, blue circles); obtained by adding the 
rotational correction $E_{\rm RC}$ on top of the PNP values (PNP$+$RC, empty 
squares). The insets show the convergence of 
the PNP and PNP\&AMP for the pre-scission configuration
with respect to 
the number of gauge angles $N_{\varphi}$ and the number of rotational angles 
$N_{\beta}$, respectively. (b) Ratio $R_{\rm RC}=(E_{\rm PNP}^{0^+}-E_{\rm PNP})/E_{\rm PY}$; 
see text for details.}
\label{fig:1D_PROJ}
\end{figure}

In the upper panel of Fig.~\ref{fig:1D_PROJ} we plot the deformation energy of 
$^{240}$Pu along the least-energy fission pathway as calculated in the HFB 
approximation (turquoise squares). $95$ configurations along the pathway 
were determined by constraining quadrupole moments within a range 
$21 \le q_{20} \le 397$ (in b) with steps $\Delta q_{20} = 4$ b, while 
$q_{30}$ moments were left unconstrained and determined self-consistently. We 
then projected these configurations onto good values of particle numbers (PNP, 
red triangles) and onto good values of particle numbers, angular momentum ($J=0$), 
and parity (PNP\&AMP, blue circles). The two insets in the upper panel of 
Fig.~\ref{fig:1D_PROJ} show the convergence of the PNP and PNP\&AMP procedures 
with respect to the number of integration points for the pre-scission 
configuration, $(q_{20}, q_{30}) = (345.0\;\text{b}, 42.6\;\text{b}^{3/2})$, 
where the underlying basis is the most incomplete.
In order to ensure proper numerical convergence across all considered 
configurations, we set $N_{\varphi} = 9$ and $26 \le N_{\beta} \le 30$. 

%As is well known from fission theorists, 
The potential energy curve of 
$^{240}$Pu is characterized by two minima, the ground state and a fission 
isomer, and two fission barriers \cite{bjornholm1980}. The scission point is 
marked by a sharp drop in energy, which occurs here at $q_{20} \approx 345$ b. 
%The five vertical lines in Fig.~\ref{fig:1D_PROJ} mark the most relevant 
%configurations along the pathway and 
Table \ref{tab:energies} lists the 
corresponding energies of these configurations. Although the HFB energy of the inner barrier ($9.37$ 
MeV) is about $3.3$ MeV higher than the empirical value inferred from fission 
cross sections \cite{smirenkin1993,capote2009}, this is mostly caused by the 
omission of triaxial effects in our calculations \cite{larsson1972,girod1983}: 
including them lowers the height of the first barrier by about $1.7$ 
MeV \cite{schunck2014}. This effect is amplified by symmetry restoration, which 
lowers the barrier by an additional $1.3$ MeV, pushing the theoretical value 
well within the uncertainty limits of the empirical value (typically about $1$ 
MeV). The outer barrier is axially symmetric and reflection asymmetric 
\cite{schunck2014}. Its height is lowered by as much as $2.3$ MeV by the 
symmetry restoration, again pushing the theoretical value within the $1$ MeV 
limit of the empirical value, $5.15$ MeV \cite{capote2009}. On the other hand, 
the HFB energy of the fission isomer is already in decent agreement with the 
empirical value of $2.25 \pm 0.20$ MeV \cite{hunyadi2001}: symmetry restoration 
degrades this agreement. These numbers are consistent with those reported in~\cite{bender2004a}.

\begin{table}[!htb]
\caption{\label{tab:energies} Calculated excitation energies of the inner 
barrier, fission isomer, and outer barrier configuration, as well as 
the pre-scission energy (in MeV) along the HFB least-energy pathway, obtained 
with HFB, PNP, and PNP\&AMP models.}
\begin{ruledtabular}
\begin{tabular}{lccc} 
Configuration  & HFB & PNP & PNP\&AMP \\
Inner barrier  &     9.37   &     8.78   &     8.05       \\ 
Fission isomer &     2.67   &     2.27   &     1.02       \\
Outer barrier  &     6.75   &     6.28   &     4.58       \\
Pre-scission   &    11.68   &    10.88   &    11.85       \\
\end{tabular}
\end{ruledtabular}
\end{table}

While previous work in Refs.~\cite{bender2004a,samyn2005} was exclusively 
focused on the potential energy curve near the two barriers, we extend this 
study all the way to the scission point. Of particular interest is the 
pre-scission energy, which is defined as the energy difference between the outer 
barrier and the scission configuration, and which may provide an important 
contribution to the excitation energy of fission fragments. Interestingly, even 
though the corrections to the barriers are significant, we find that the total 
correlation energy beyond the outer barrier saturates, with the result that 
symmetry restoration has a negligible impact on the value of pre-scission 
energy. 

In many studies of spontaneous fission, the effect of AMP is simulated by 
what is known as the rotational energy correction \cite{schunck2016,egido2004}. It was 
observed in \cite{egido2000} that this term is well approximated by 
$E_{RC} = 0.7 \times E_{PY}$ where  
$E_{PY} = - \langle \mathbf{J}^2\rangle/(2\mathcal{J}_{PY}$),  
$\langle\mathbf{J}^2\rangle$ is the total angular momentum dispersion,
$\mathcal{J}_{PY}$ is the Peierls-Yoccoz moment of inertia \cite{peierls1957}, and the phenomenological 
quenching factor $0.7$ is included to account for approximations introduced in 
calculating $\mathcal{J}_{PY}$. In panel (a) of Fig.~\ref{fig:1D_PROJ} we 
also show the curve obtained by adding $E_{\rm RC}$ on top of the calculated PNP 
values (PNP$+$RC), while the ratio 
$R_{\rm RC} = (E_{\rm PNP}^{0^+}-E_{\rm PNP})/E_{\rm PY}$
%between the calculated rotational energy, 
%$E_{\rm PNP}^{0^+}-E_{\rm PNP}$, and $E_{\rm PY}$ values 
is shown in the lower panel of 
Fig.~\ref{fig:1D_PROJ}. Our calculations 
confirm that $E_{\rm RC}$ is an excellent approximation to the 
exact model at very large deformation and all the way to the scission point.
However, we also observe that for configurations with 
$q_{20} \lesssim 150$ b the quenching factor of $0.7$ is not sufficient, 
leading to differences in energy up to $2.5$ MeV. 
This discrepancy could have a severe impact on observables that are very 
sensitive to details of the underlying PES, such as the spontaneous fission 
half-lives \cite{warda2012,baran2015}.

{\it 2D PES and fission fragment distributions. } While 1D fission 
paths can be sufficient to compute observables such as half-lives and cross 
sections, quantities such as fission fragment distributions require probing at 
least two dimensions in the collective space. Starting from the PES of 
$^{240}$Pu in the HFB approximation reported in \cite{schunck2014}, we thus 
selected a total of $1150$ configurations within 20 MeV of the ground state 
energy. They cover a very broad range of quadrupole and octupole deformations, 
with $20 \le q_{20} \le 567$ (in b) and  $0\le q_{30} \le 70$ (in b$^{3/2}$). 
We then projected each of these configurations onto good values of particle numbers, 
angular momentum, and parity using the above method \cite{supplemental_material}.

\begin{figure}[!htb]
\includegraphics[width=0.49\textwidth]{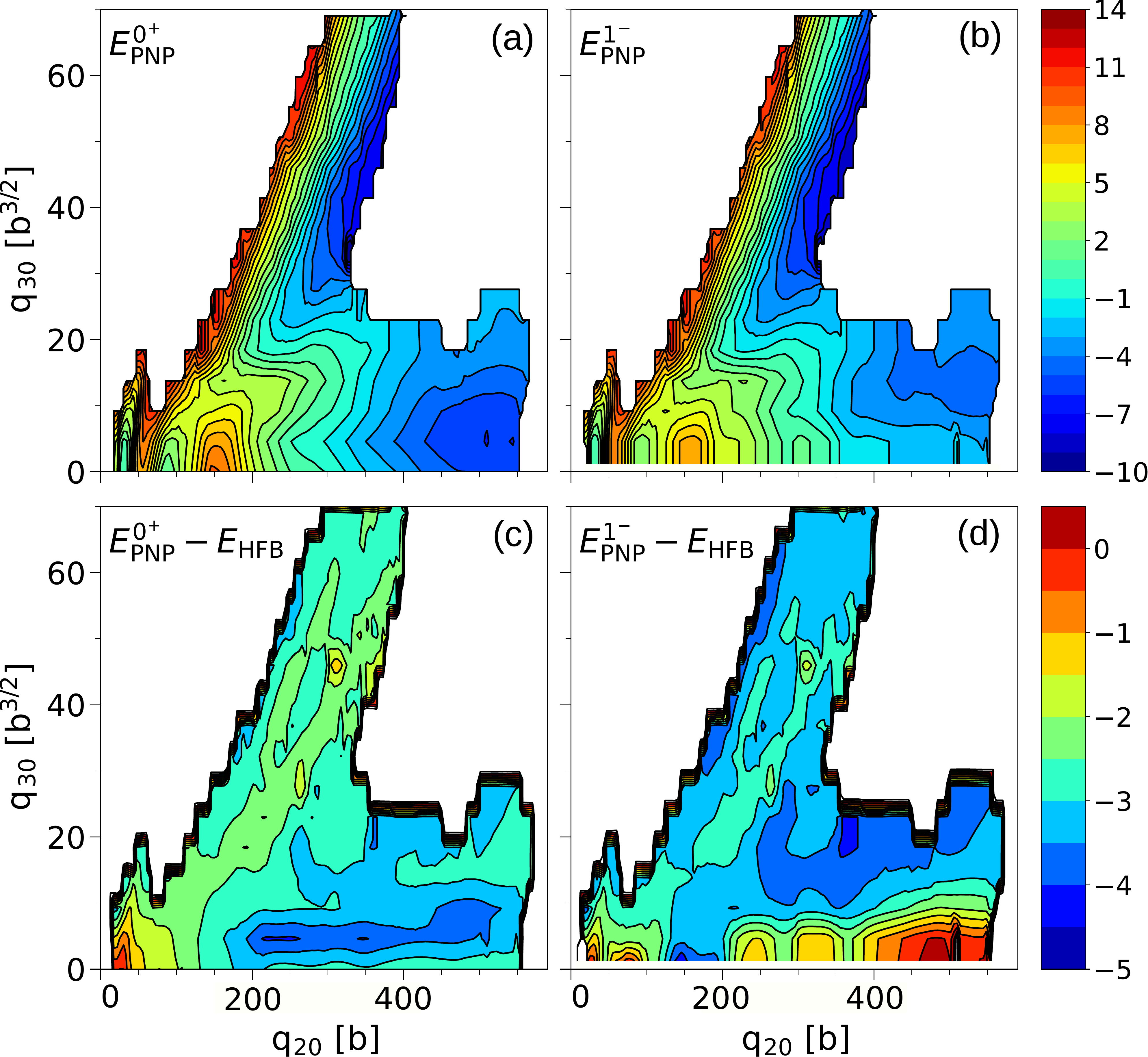}
\caption{(a) Two-dimensional symmetry-restored PES of 
$^{240}$Pu in the $(q_{20}, q_{30})$ plane for the $J^{\pi}=0^+$ configuration.
(b) Same for $J^{\pi} = 1^-$. Both surfaces are 
normalized with respect to the energy of their respective ground states. 
(c) Difference between the symmetry-restored and the HFB surface for the 
$J^{\pi}=0^+$ configuration. (d) Same for the $J^{\pi} = 1^-$.}
\label{fig:2D_PROJ}
\end{figure}

Figure \ref{fig:2D_PROJ} shows the PES for the $J^{\pi} = 0^{+}$ 
(a) and $J^{\pi} = 1^{-}$ (b) states with the exact number 
of particles $N = 146$ and $Z = 94$. Although projection on $1^{-}$ is not possible for $q_{30}=0$ b$^{3/2}$ configurations (indicated by a white line on the surface), we emphasize that the energy remains well defined and finite in the $q_{30} \rightarrow 0$ limit \cite{egido1991}. Overall, the PES retains its main 
features such as the fission isomer and the main fission valley, which extends 
from $(q_{20},q_{30}) \approx (90\; \text{b},0\; \text{b}^{3/2})$ to 
$(q_{20},q_{30}) \approx (340\; \text{b},40\; \text{b}^{3/2})$. Panels (c) and (d), 
which show the energy difference between the symmetry-restored and HFB 
surfaces, provide an additional insight. In particular, for the $0^{+}$ 
state, a pronounced gain in energy is observed at low $q_{30}$ values for a 
wide range of configurations, pointing to the possible enhancement of symmetric fission.
%, which could enhance the symmetric fission $(A=120)$. 
For the $1^{-}$ state, the correlation energy is large along and around the 
least-energy pathway, suggesting broader fission fragment distributions. 
%We note that significant lowering 
%of configurations in the $(q_{20}, q_{30}) \approx (350\;\text{b}, 20\;
%\text{b}^{3/2})$ region could result in favoring less asymmetric fission 
%fragmentations, particularly in the $1^{-}$ case.

\begin{figure}[!htb]
\includegraphics[width=0.49\textwidth]{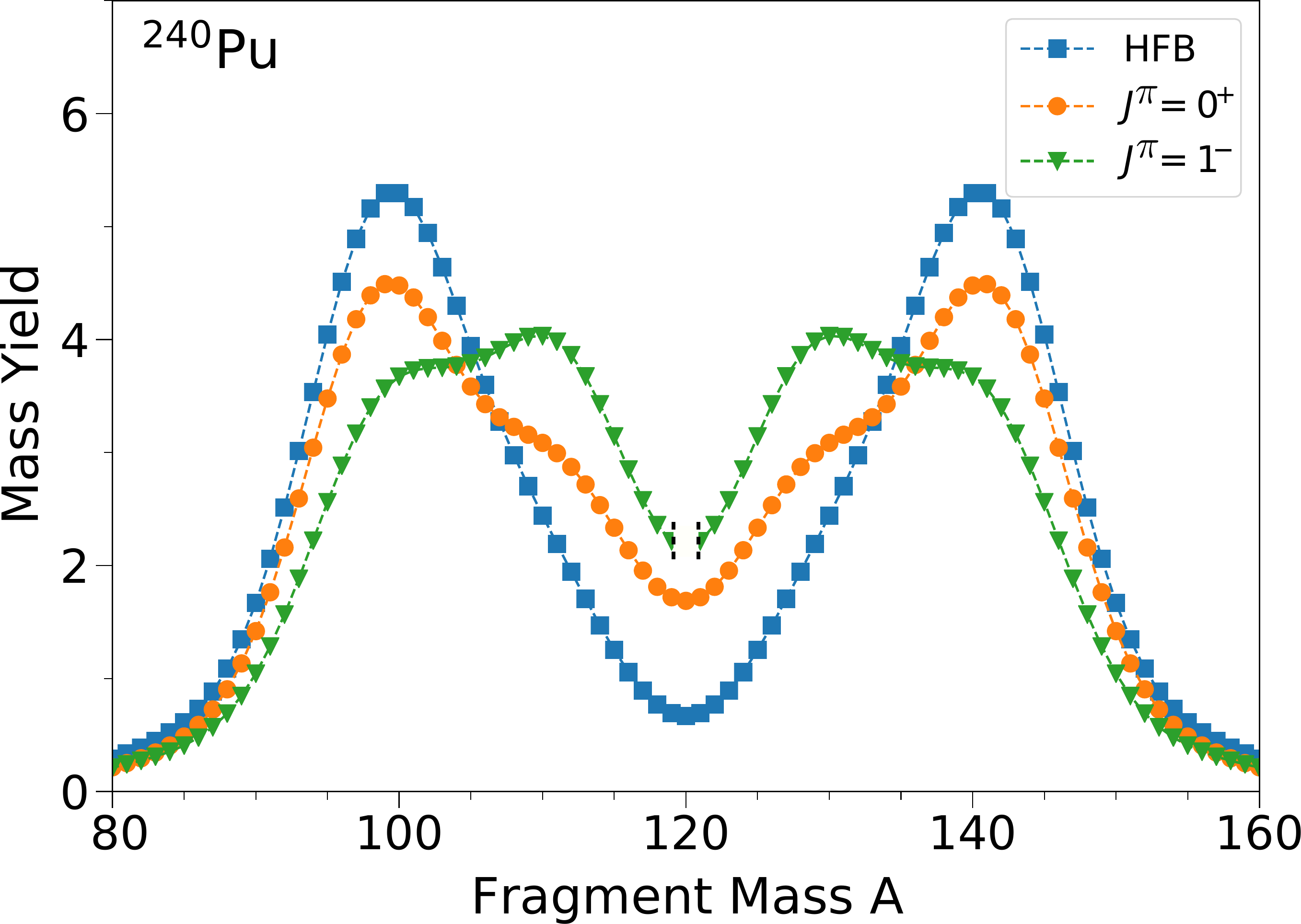}
\caption{Primary fission fragment mass distributions in $^{240}$Pu obtained 
from FELIX with the PES at the HFB level (blue squares) or after PNP\&AMP for 
the $0^{+}$ (orange circles) or $1^{-}$ (green triangles) states. }
\label{fig:yields}
\end{figure}

To estimate the actual effect on fission fragment distributions \cite{supplemental_material}, we used the 
FELIX solver \cite{regnier2018a} to solve the collective Schr\"odinger equation 
originating from the Gaussian overlap approximation of the time-dependent GCM. The 
inputs to FELIX were the PESs (HFB and PNP\&AMP for the 
$0^{+}$ and $1^{-}$ states), the GCM inertia tensor computed at the 
perturbative cranking approximation, and scission configurations defined by the 
HFB expectation value of the Gaussian neck operator $q_{\rm N} = 5$ with a 
folding factor of width $\sigma_{A} = 5$; see \cite{regnier2016a} for a 
discussion. To simulate the neutron-induced fission for thermal neutrons, the energy of the initial state was set at $1$ MeV above the inner barrier.
Figure \ref{fig:yields} demonstrates the impact on the fragment mass distributions:
%the primary fission fragment mass distributions in $^{240}$Pu obtained 
%with HFB and PNP\&AMP ($J^{\pi}=0^+$ and $J^{\pi}=1^-$) surfaces as microscopic 
%inputs. 
fission becomes more symmetric after 
projections, and the symmetric fission mode is indeed enhanced for the $0^+$ 
state. Furthermore, the distribution for the $1^-$ state is significantly broadened 
and favors less asymmetric fragmentations. Note that yields stemming from $q_{30} = 0$ b$^{3/2}$ and nearby configurations are cautiously excluded from the plot (indicated by a gap in the 
curve). These results represent the first attempt to quantify the effect of symmetry 
restoration on actual fission observables. A fully consistent determination of 
fission fragment distributions will require developing a projected theory of 
collective inertia, estimating the spin distribution of the fissioning nucleus, and 
generating PESs with a much higher resolution in $(q_{20},q_{30})$.

{\it Conclusion.} Restoring broken symmetries is a necessary step for nuclear 
models to describe different fission channels. In this work, we reported the 
first symmetry-restoring description of fission from the ground state to scission. Our analysis of the benchmark case of 
$^{240}$Pu indicates that projection correlation energies cannot be 
approximated by a phenomenological formula across the entire range of 
deformations relevant for fission, and that symmetry restoration may have a 
substantial impact on the mass distribution of fission fragments. These 
conclusions should be validated by developing a projected theory of collective 
inertia. The technique of symmetry restoration in incomplete bases is extendable to configuration mixing schemes, and could therefore be key to providing a
reliable and computationally feasible framework for nuclear structure studies
relevant to ongoing experimental programs at radioactive beam facilities.

%%%%%%%%%%%%%%%%%%%%%%%%%%%%%%%%%%%%%%%%%%%%%%%%%%%%%%%%%%%%%%%%%%%%%%%%%%%%%%%
%%%%%%%%%%%%%%%%%%%%%%%%%%%%%%%%%%%%%%%%%%%%%%%%%%%%%%%%%%%%%%%%%%%%%%%%%%%%%%%
%%%%%%%%%%%%%%%%%%%%%%%%%%%%%%%%%%%%%%%%%%%%%%%%%%%%%%%%%%%%%%%%%%%%%%%%%%%%%%%
%%%%%%%%%%%%%%%%%%%%%%%%%%%%%%%%%%%%%%%%%%%%%%%%%%%%%%%%%%%%%%%%%%%%%%%%%%%%%%%

We thank M. Verri\`ere for many fruitful discussions and R. Vogt for careful reading of the 
manuscript. This work was performed under the auspices of the U.S.\ 
Department of Energy by Lawrence Livermore National Laboratory under Contract 
DE-AC52-07NA27344 (NS). Computing support for this work came from the Lawrence 
Livermore National Laboratory (LLNL) Institutional Computing Grand Challenge
program.

%%%%%%%%%%%%%%%%%%%%%%%%%%%%%%%%%%%%%%%%%%%%%%%%%%%%%%%%%%%%%%%%%%%%%%%%%%%%%%%
%%%%%%%%%%%%%%%%%%%%%%%%%%%%%%%%%%%%%%%%%%%%%%%%%%%%%%%%%%%%%%%%%%%%%%%%%%%%%%%
%%%%%%%%%%%%%%%%%%%%%%%%%%%%%%%%%%%%%%%%%%%%%%%%%%%%%%%%%%%%%%%%%%%%%%%%%%%%%%%
%%%%%%%%%%%%%%%%%%%%%%%%%%%%%%%%%%%%%%%%%%%%%%%%%%%%%%%%%%%%%%%%%%%%%%%%%%%%%%%

\bibliographystyle{unsrt}
%\bibliography{zotero_output,books}
\bibliography{Pu240Letter}

\end{document}